\documentstyle[twocolumn,prc,aps,epsfig]{revtex}

\begin{document}
\newcommand{\be}{\begin{eqnarray}}
\newcommand{\ee}{\end{eqnarray}}

\twocolumn[\hsize\textwidth\columnwidth\hsize\csname @twocolumnfalse\endcsname
\title{The Azimuthal Asymmetry at large $p_t$ \\
seem to be too large for a pure ``Jet Quenching''}
\author{E. V. Shuryak}
\address{Department of Physics and Astronomy, State University of New York,
Stony Brook, NY  11794-3800}

\date{\today}
\maketitle
\begin{abstract}
We discuss simple generic model of ``jet quenching'' in which
matter absorption is defined by one parameter. We show that as absorption
grows, the azimuthal asymmetry parameter $v_2$ grows as well, reaching
the finite limit $v_2^*$ which has
 a simple geometric interpretation. We show 
that this limit is still below the experimental
values for $6> p_t >  2 \, GeV$, according to preliminary data from
STAR
experiment at RHIC. We thus conclude that ``jet quenching'' models
 alone cannot account for the observed phenomenon, and speculate about 
alternative scenarios.
\end{abstract}
\vspace{0.1in}
]

\begin{narrowtext}   
\newpage

1.Azimuthal asymmetry for non-central heavy ion collisions
have been predicted \cite{QM99} to be larger at RHIC then at lower 
energies.
In hydrodynamic models this happens due to the stronger  push
by high pressure of Quark-Gluon Plasma well above the phase
transition region, which is 
expected to be produced at RHIC. In contrast to that,  models
based on string picture of hadron production (e.g. RQMD and UrQMD
event generators)
or on
mini-jet scenarios (e.g. HIJING) have predicted its decrease. 
The issue has been settled already by the first data from RHIC,
by STAR collaboration \cite{STAR}, which have found large asymmetry
consistent with hydrodynamic predictions. Detailed studies 
\cite{Kolb,TLS}
have provided
significant details, such as the asymmetry parameter
\be v_2=<cos(2\phi)>\ee
(where $\phi$ is the angle between the impact parameter and
momentum of a secondary hadron in the
 transverse plane) as a function of centrality, particle
type and its $p_t$. Data from STAR and PHENIX
experiments  for rather wide range of momenta $p_t< 2 \, GeV$  agree
well with these predictions for all secondaries. 
This is probably the most direct
signature of QGP plasma formation, observed at RHIC.

In this letter we however discuss a different question, related with
(much less certain experimentally) such asymmetry 
at $higher$ transverse momenta  $6> p_t >  2 \, GeV$. 
According to the latest  STAR data \cite{STAR_higpt} (which
are still unpublished and are thus considered preliminary, although
reported at many meetings), $v_2(p_t)$ for all charged secondaries 
seem to be about constant, for each centrality. This means
 a different regime seem to be
established in this region of $p_t$, and
the original intention of this note was to compare these data
with a simple geometric model for jet quenching by relating the
asymmetry to the strength of the jet quenching itself. However, 
after playing with different
versions of the model, from more complex to a most generic one
to be reported in this note,
 I concluded that the intended fit is simply impossible.

A ``jet quenching'' idea has been discussed for a long
time, see e.g.
 \cite{WG}, and it has been naturally related to the azimuthal asymmetry
 for non-central collisions. 
If a high-$p_t$ jet is loosing energy in matter, jet emission is
dominated by the surface of the almond and the correlation
between position and the emission direction appears, thus the
observed azimuthal asymmetry.

A relation between this phenomenon and data has been discussed in
\cite{GVW},
where it was concluded that a $combination$ of jet quenching and
hydrodynamical expansion 
 can approximately describe them\footnote{ Although no curves for jet 
quenching alone are shown, the text  implies that it is indeed insufficient
by itself, in agreement with the (more general) argument we will give
below.}. 
 Later STAR data
 have  shown at high  $p_t=2-6$ an approximately $p_t$-independent
$v_2$, which disagree  with a decreasing trend expected from jet 
quenching. Qualitative discussion of many possible scenarios which can 
have such a behavior has been 
made in Ref.\cite{GVWH}, including the interplay of jet quenching,
hydrodynamical expansion and ``baryon junction dynamics''.  We return
to this discussion at the end of the paper.

2.The present work ignores such details as $p_t$ dependence of
$v_2$ and focuses instead on its measured $values$:  we demonstrate that looking at pure
geometric aspect of the problem one can show that those are too high for 
$any$ jet quenching model (without hydro).

The
 most generic model we use can be described as follows.
First, the
distribution
of origination points for outgoing jets is simulated: 
this is done using the usual assumption
of parton model and the simplest model of nuclei as two
homogeneous colliding spheres. (Diffuse boundary only makes effects smaller.)

The second step is the calculation of
the chances for the parton to escape the absorption
in matter, as it goes out of the almond. The absorption rate 
 is characterized by one (and the only) {\em free parameter} of the model.
Its magnitude determines the strength of jet quenching itself
(the fraction of escaping partons $f(p_t,b)$), with the predicted
azimuthal asymmetry, $v_2(b)$.

As high-$p_t$ partons move with the speed of light, we ignore
possible change of shape due to geometrical
expansion of the ``almond'' during this time. (If anything, this will
reduced the asymmetry, as expansion reduces spatial asymmetry.)

Naturally, 
in the absence of an absorption there is no azimuthal asymmetry,
$v_2(\kappa=0)=0$, while increasing absorption creates increasing $v_2$. Interestingly,
in the limit of very strong absorption  the asymmetry
 reaches a {\em finite limit},  denoted by asterix below
\be v_2(b,\kappa\rightarrow\infty)\, \rightarrow \, v_2^*(b) \ee
The reason for that is that in this case all the
emitted partons/hadrons originate from the thin surface of the almond
(see below). Even in this case, however, partons have
 half solid angle open for them: thus $v_2^*(b)$
 has  direct geometric interpretation. The main point of this letter
 is that, after evaluating  $v_2(b)$ values for the experimental
conditions and comparing it with data we have found that even the
limiting ones,   $v_2^*(b)$, are below the data.

\begin{figure}[h]
%   \vskip 0.2in
\begin{center}
   \includegraphics[width=2.in]{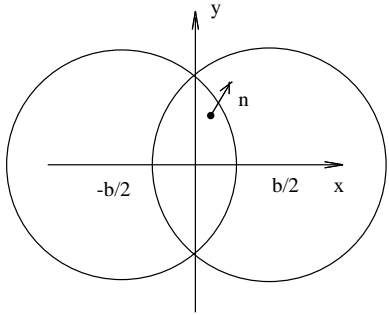}
\end{center}
%   \vskip 0.2in
   \caption[]{A frontal view of two colliding nuclei, with
   definition of the axis. The black dot (x,y) inside the almond is
the origin of the parton, which propagates in the direction of the
unit vector $\vec n$. }
\end{figure}
%======================================================================

3.Let us now provide more details about the model itself.
In fig.1 we show geometry of the collision and definition of two 
longitudinal lengths, $L_\pm(x,y)$ for a hard 
collision at point  (x,y). For hard spheres
\be 
L_\pm(x,y)=2[R^2-y^2-(x\pm b/2)^2]^{1/2}
\ee
The probability of production of a parton
in hard collisions at position x,y is simply
proportional to the product of longitudinal lengths $ P(x,y)=\alpha
L_+(x,y) L_-(x,y)$. Vanishing of each
of these factors defines the boundary of the initial almond in the
transverse plane. Fig.1
shows the sketch of the initial distribution in transverse, x-y, plane.
One characterizes it by the standard spatial anisotropy 
\be s_2(b)=<y^2-x^2>/<y^2+x^2> \ee
where angular brackets means average over  all produced jets, with the
weight given by the parton model as described above. The distribution
depends on impact parameter b, indicated in the l.h.s. In the table
below we will make integration over b with geometric weight $2\pi b
db$ over bins of centrality,
within limits defined by upper and lower percentage of the total cross 
section.

The probability to escape depends not only on the point of jet origin
but also
on the optical depth of matter along the
outgoing line, $(x+s*n_x,y+s*n_y)$ 
which we calculate as follows
\be 
f=exp[- \kappa \int_0^\infty ds (L_- L_+)(x+s*n_x,y+s*n_y)]
\ee
The parameter $\kappa$ (dimension $fm^{-3}$)
includes both the density of the material and the absorption rate.
The following fig.2 shows how the efficiency of the
parton quenching and the $v_2$ parameter depend on
it,
in the whole dynamical range.
The dependence of jet quenching and $v_2$ on the absorption strength
is shown in Fig.2(b). It displays the saturation of $v_2$ 
as well as the tendency toward the surface emission at large absorption,
mentioned in
the introduction.

The main outcome of the simulations is summarized 
in the Table 1, in which we compare the high absorption limit
$v_2$ calculated from the model with STAR preliminary
data\footnote{The error bars are calculated by the author, based on
 three  STAR points at the 
largest $p_t$ bins, for 
each centrality. As at this point the data still have preliminary
status, the reader should be warned that
the error bars may be modified and the systematic
errors be better understood and included.  Now it is not possible
to quantify the problem we discuss any better.}
\cite{STAR_higpt}
at $p_t>3.5 
\, GeV$.  As one can easily see, even in the high absorption limit
the model {\em fails to reproduce data}, being systematically below
the 
present preliminary data. The difference is especially striking for the
most central bin, in which the observed $v_2$ nearly matches
the asymmetry $s_2$ of the original almond.

(Although the result is described as that in the
 high absorption limit, it actually
corresponds to the calculation in which absorption was large but finite,
 $\kappa=.2\,fm^{-3}$, with 
 the actual quenching factors $f$ also given in the table.)

\begin{table}[h]

\begin{tabular}{ccccc}
 Centrality \% & $<f>$ & $v_2^*/s_2$ & $v_2^*$ & $v_2^{STAR}$ \\
 0-11 & .018 & .32  & .042 & .12$\pm$ 0.02 \\
 11-34 & .027  & .35  & 0.12 & .16$\pm$ 0.02 \\
 34-85 & .046 & .31  & 0.16  & .22$\pm$ 0.02 \\
\end{tabular}
\caption{The limiting momentum/spatial asymmetry for three different
centrality selections of STAR, given as $v_2$ versus the
percentage of total AuAu cross section. The quantity
 $<f>$ is the  escape probability  (5) averaged over  produced
jets in the collisions,  with all directions and origin points. }
\label{tab_parts}
\end{table}

\begin{figure}[h]
%   \vskip 0.2in
\begin{center}
   \includegraphics[width=3.in]{f_kappa.eps}
   \vskip 0.2in
   \includegraphics[width=3.in]{v2_kappa.eps}
%   \vskip 0.2in
   \includegraphics[width=3.in]{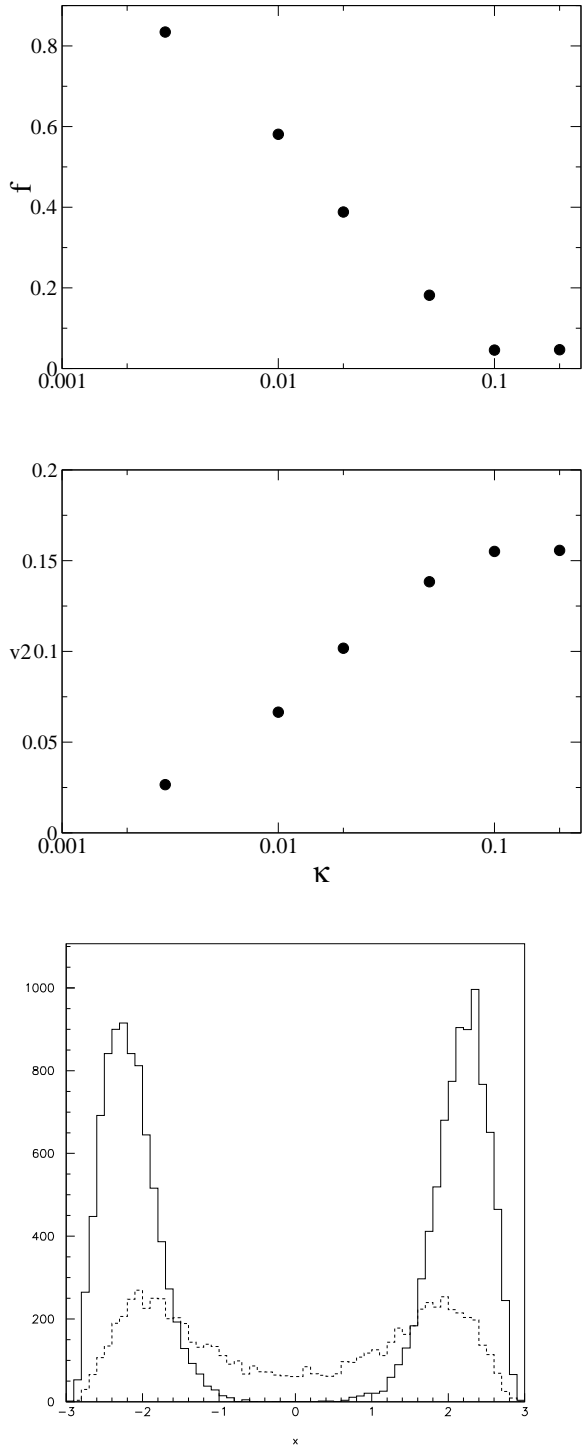}
\end{center}
%   \vskip 0.2in
   \caption[]{
   Dependence of (a) the strength of ``jet quenching'', the fraction
of escaping partons $f(\kappa)$ and (b) the corresponding asymmetry parameter $v_2(\kappa)$ 
 on the absorption strength parameter $\kappa$, in $fm^{-3}$ .
Figure (c) show the (un-normalized) 
distribution over initial position x of the
escaped partons, for the cut $|y|< 2 \ fm$. The solid and dashed lines 
are for $\kappa=0.2, 0.05 \, fm^{-3}$ respectively.
 All three figures correspond to  the most peripheral centrality
bin, 34-81 percent. }
\end{figure}
%======================================================================

5.Let us now summarize the main result of this letter:
the dynamical range of ``jet quenching'' scenarios is
approximately confined in the region
\be 0 \, < \, v_2^*/s_2 \, <\, 1/3 \ee
while the  preliminary STAR data give larger values $v_2/s_2=.5-1$
and therefore they  cannot be explain by models of this kind {\em alone} , no matter
what 
magnitude of jet quenching be used.
The generic model used
 can of course be modified in many ways, but it seems unlikely that
jet quenching  by {\em matter absorption} in whatever form is
able to explain these  data by itself. 

{\em Assuming these preliminary STAR  data are correct},
let us consider
what  their explanation can be.
The main shortcoming of the model comes from the idea that secondaries
in this region of $p_t$ originate $only$ from  jets, obtaining
azimuthal asymmetry $only$ from
geometrical asymmetry of the almond. The interplay between jet quenching
and hydro expansion, quantitatively discussed in \cite{GVWH}, only reduces 
the effect due to reduction of geometrical asymmetry with time.

The resolution of this puzzle can only be obtained if 
 a significant fraction of
 secondaries  originate from a  {\em source other than jets}.
A general discussion  in \cite{GVWH} have mentioned a possibility that
$v_2$ for baryons and pions can be very different, with the former
getting a contribution from
 {\em ``baryon
junctions''} and/or {\em collective flow}, as  the sources complementary to jets. 
We also think that it is  likely to be the explanation, although
we are quite sceptical about the role of the
 baryon junctions\footnote{
Recent STAR data on spectra of $\phi$ mesons have provided one more
argument against it. These data
show that $\phi$  has $p_t$ 
slopes consistent with hydro predictions \cite{TLS} with the slope not very different
from the nucleon's: so it is the mass not the baryon number which
matters
here.}.

Collective hydro expansion is not just  a simple 
and general concept,  it   
  is  basically the only known
mechanism  capable to generate very large values of the
azimuthal anisotropy. (Let us remind the reader why is it so.  Due to different hydro
motion in different
directions, spectra have the $\phi$-dependent  $p_t$ $slopes$,
resulting in asymmetry, $v_2$, about linearly increasing with $p_t$ .)
However, the issue is far from being simple, and
 a significant role of hydro component 
in the high-$p_t$ tails of spectra, at
  $p_t\sim 4-6 \, GeV$, is a very non-trivial thing. These 
 tails of the particle spectra are  6 orders of magnitude
below the majority of the particles, 
 way below where a macroscopic language
is routinely  used. 
More work is needed in order to understand whether such approaches can 
at all be used in this region. In connection with that let me mention
a very interesting paper by Molnar and Gyulassy
\cite{Molnar} in which
$v_2$  has been generated  kinetically in some model with very large
cross section, way above perturbative predictions. Although
 it is far from being
clear that the extreme assumptions made in these calculations are
realistic, it has been able to yield collective flow and sufficient values of the $v_2$.

Experimentally it is quite obvious what one should do:
as soon as statistics will allow, to study $v_2$ at such $p_t$
for {\em any identified secondaries}. Particles which are seen via decays, 
e.g.
$\Lambda$ and $K_s$ and especially $\phi$ can be identified at rather high momenta and are 
thus most interesting.
 Since jets decay into pions much
more than into strange mesons like $\phi$ and  especially into
baryons, one should expect the corresponding fractions
of jet-originated and hydro-originated secondaries be very
different for all of them. 
The observed constancy of $v_2$ with $p_t$ for all charge secondaries
is likely to be just a result of occasional cancellation
between rising hydro-based  and decreasing jet-based components.

{\bf Note Added in proofs} 
After the paper was submitted to PRC,  STAR data used 
have passed necessary procedures and are no longer preliminary. The
final data are submitted to PRL for
publication, as ``Azimuthal anisotropy and correlations in the hard
scattering regime at RHIC'', by C.Adler et al,nucl-ex/0206006.
Systematic effects due to two-body correlations have been studied
by comparison between 2-particle and 4-particle cumulants.
 When the latter values for $v_2$ is used, the discrepancy with the maximal
model values at strong jet quenching $v_2^*$  nearly disappears.
Another significant fact reported in this STAR  publication is the
first direct observation of jet component in the 2-body correlations.  
More recent STAR data at 200 GeV/N have been presented
 at recent Quark Matter 2002
conference. Due to much higher statistics, those extends to larger 
$p_t=6-12 \, GeV$, but display about the same $v_2$.   
Good agreement between the measured values
of $v_2$ and the theoretical high quenching limit
 $v_2^*$  has been shown
 in the summary talk there by S.A.Voloshin: it seem to suggest
that geometrical interpretation of azimuthal asymmetry 
suggested in this paper seem to be correct, after all.

This work is partially supported by the US-DOE grants DE-FG02-88ER40388
and DE-FG03-97ER4014.

\end{narrowtext}
\end{document}